\documentclass[]{spie}  

\usepackage{amsmath,amsfonts,amssymb}
\usepackage{graphicx}
\usepackage[colorlinks=true, allcolors=blue]{hyperref}

\title{Final Alignment and Image Quality Test for the Acquisition and Guiding System of SOXS}

\author[a]{Jos\'e A. Araiza-Dur\'an}
\author[b]{Giuliano Pignata}
\author[a,c]{Anna Brucalassi}
\author[d]{Matteo Aliverti}
\author[e]{Federico Battaini}
\author[e]{Kalyan Radhakrishnan}
\author[e]{Simone Di Filippo}
\author[e]{Luigi Lessio}
\author[e]{Riccardo Claudi}
\author[e]{Davide Ricci}
\author[f]{Mirko Colapietro}
\author[i,j]{Rosario Cosentino}
\author[f]{Sergio D'Orsi}
\author[j]{Matteo Munari}
\author[e]{Marco Dima}
\author[f]{Pietro Schipani}
\author[d]{Sergio Campana}
\author[e]{Andrea Baruffolo}
\author[f]{Ricardo Zanmar Sanchez}
\author[d]{Marco Riva}
\author[d]{Matteo Genoni}
\author[g,h]{Sagi Ben-Ami}
\author[n]{Adam Rubin}
\author[g]{Rachel Bruch}
\author[f]{Giulio Capasso}
\author[k]{Francesco D'Alessio}
\author[d]{Paolo D'Avanzo}
\author[g]{Ofir Hershko}
\author[l,m]{Hanindyo Kuncarayakti}
\author[d]{Marco Landoni}
\author[o,j]{Salvatore Scuderi}
\author[k]{Fabrizio Vitali}
\author[p]{David Young}
\author[q]{Jani Achrén}
\author[r]{Iair Arcavi}
\author[e]{Enrico Cappellaro}
\author[f]{Massimo Della Valle}
\author[j]{Rosario Di Benedetto}
\author[g]{Avishay Gal-Yam}
\author[i]{Marcos Hernandez Díaz}
\author[m,l]{Jari Kotilainen}
\author[s]{Gianluca Li Causi}
\author[f]{Laurent Marty}
\author[l]{Seppo Mattila}
\author[g]{Michael Rappaport}
\author[e]{Bernardo Salasnich}
\author[u,p]{Stephen Smartt}
\author[t]{Maximilian Stritzinger}
\author[i]{Hector Pérez Ventura}
\author[d]{Laura Asquini}
\author[g]{Alex Bichkovsky}
\author[f]{Salvatore Savarese}
\author[e]{Lorenzo Cabona}

\affil[a]{INAF - Osservatorio Astronomico di Arcetri, Via E. Fermi 5 Firenze, Italy}
\affil[b]{Universidad de Tarapacá, Casilla 7D, Arica, Chile}
\affil[c]{Universidad Andrés Bello, Avda. República 252, Santiago, Chile}
\affil[d]{INAF - Osservatorio Astronomico di Brera, Via Bianchi 46, I-23807 Merate, Italy}
\affil[e]{INAF - Osservatorio Astronomico di Padova, Vicolo dell’Osservatorio 5, I-35122 Padova, Italy}
\affil[f]{INAF - Osservatorio Astronomico di Capodimonte, Salita Moiariello 16, I-80131 Napoli, Italy}
\affil[g]{Weizmann Institute of Science, Herzl St 234, Rehovot, 7610001, Israel}
\affil[h]{Harvard-Smithsonian Center for Astrophysics, Cambridge, USA}
\affil[i]{INAF - Fundación Galileo Galilei, Rambla J.A. Fernández Pérez 7, E-38712 Breña Baja (TF), Spain}
\affil[j]{INAF - Osservatorio Astrofisico di Catania, Via S. Sofia 78, I-95123 Catania, Italy}
\affil[k]{INAF - Osservatorio Astronomico di Roma, Via Frascati 33, I-00078 Monte Porzio Catone, Rome, Italy}
\affil[l]{Tuorla Observatory, Department of Physics and Astronomy, University of Turku, FI-20014 University of Turku, Turku, Finland}
\affil[m]{FINCA - Finnish Centre for Astronomy with ESO, Turku, Finland}
\affil[n]{European Southern Observatory, Karl Schwarzschild Strasse 2, D-85748, Garching bei München, Germany}
\affil[o]{INAF - Istituto di Astrofisica Spaziale e Fisica Cosmica, Via Corti 12, I-20133 Milano, Italy}
\affil[p]{Queen's University Belfast, School of Mathematics and Physics, Belfast, BT7 1NN, UK}
\affil[q]{Incident Angle Oy, Capsiankatu 4 A 29, FI-20320 Turku, Finland}
\affil[r]{Tel Aviv University, Tel Aviv, Israel}
\affil[s]{INAF - Istituto di Astrofisica e Planetologia Spaziali, Via Fosso del Cavaliere, I-00133 Roma, Italy}
\affil[t]{Aarhus University,Ny Munkegade 120, D-8000 Aarhus, Denmark}
\affil[u]{University of Oxford,Keble Road, Oxford OX1 3RH, Oxford, UK}

\authorinfo{Further author information: (Send correspondence to J.A)\\J.A.: E-mail: jose.araiza@inaf.it}

\begin{document} 
\maketitle

\begin{abstract}

SOXS (Son Of X-Shooter) will be the new medium-resolution (R~4500 for 1” slit), high-efficiency, wide-band spectrograph for the ESO NTT at La Silla Observatory, Chile. It will be dedicated to the follow-up of any kind of transient events, ensuring fast time, high efficiency, and availability. It consists of a central structure (common path) that supports two spectrographs optimized for the UV-Visible and a Near-Infrared range. Attached to the common path is the Acquisition and Guiding Camera system (AC), equipped with a filter wheel that can provide science-grade imaging and moderate high-speed photometry. The AC Unit was integrated and aligned during the summer months of 2022 and has since been mounted in the NTT’s telescope simulator. This work gives an update on the Acquisition Camera Unit status, describes the Image Quality Tests that were performed, and discusses the AC Optical Performance.

\end{abstract}


\section{INTRODUCTION}
\label{sec:intro}

The Son Of X-Shooter (SOXS), is a spectroscopic facility for the European Southern Observatory (ESO) New Technology Telescope (NTT) at La Silla Observatory, Chile \cite{SOXS1:SPIE2018,SOXS2:SPIE2018,SOXSB:SPIE2018,SOXS9:SPIE2020,SOXSA:SPIE2020,SOXSD:SPIE2020,SOXS1:SPIE2022,SOXS9:SPIE2022,SOXSC:SPIE2022,SOXS7:SPIE2022}.
It will be dedicated to the follow-up of transient sources, proposed by an Italian led consortium.
It will be a two arm spectrograph able to cover the optical/NIR band ($0.35-1.8 \mu m$).
SOXS will be a medium resolution spectrometer ($R \sim 4500$) with light imaging capabilities in the optical with multi-band photometry for faint transients.
The spectrograph is divided into five sub-systems: Common Path (CP) \cite{SOXS3:SPIE2018,SOXS4:SPIE2020}, UV-VIS spectrograph \cite{SOXS5:SPIE2018,SOXS9:SPIE2018,SOXSB:SPIE2020,SOXS5:SPIE2020,SOXSA:SPIE2022}, NIR\cite{SOXS6:SPIE2020}, and the Acquisition Camera (AC) \cite{SOXSA:SPIE2018,SOXS2:SPIE2020,SOXS8:SPIE2022}. 

\begin{figure} [ht]
\begin{center}
\begin{tabular}{c} 
\includegraphics[height=7.5cm]{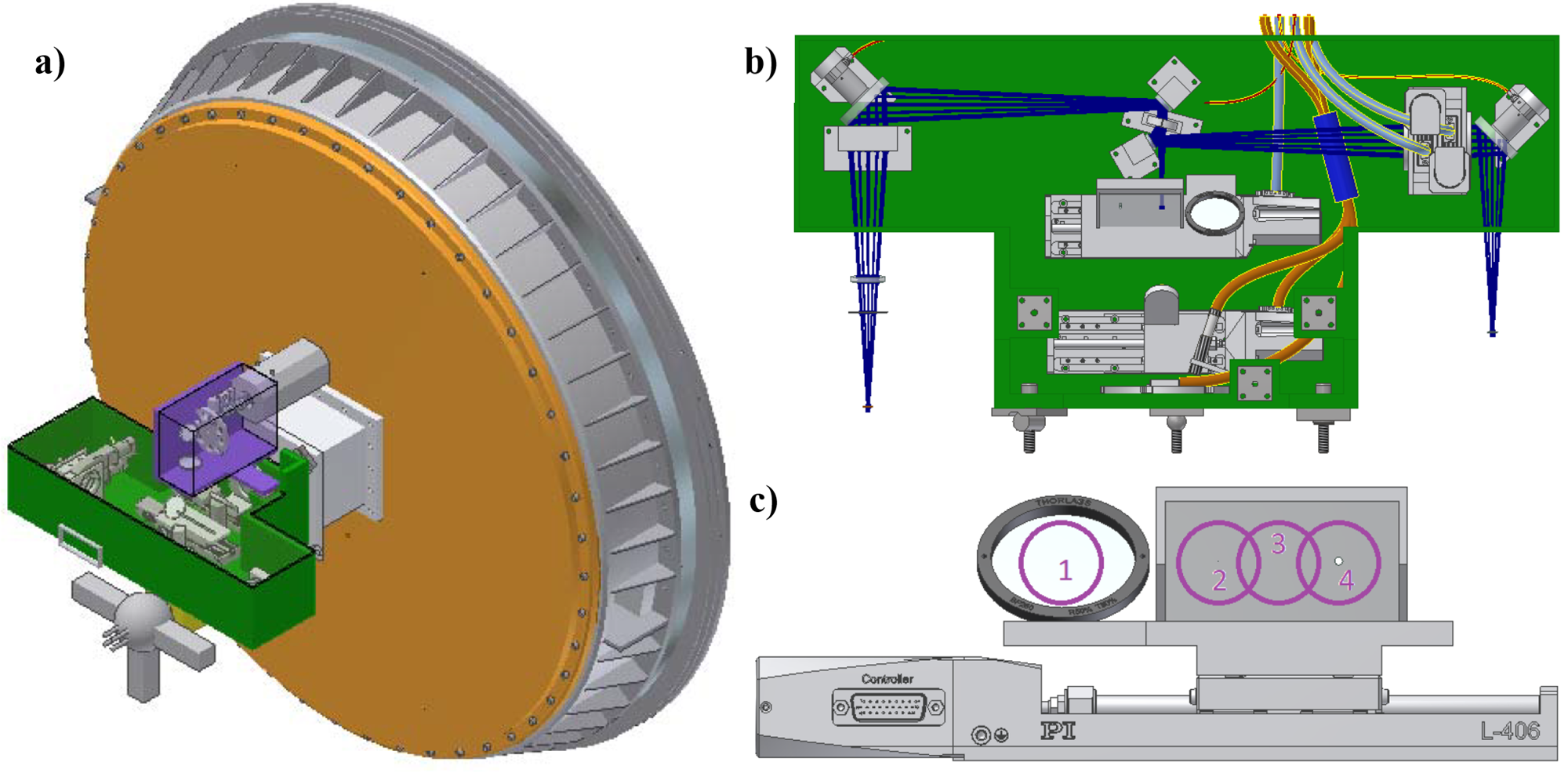}
\end{tabular}
\end{center}
\caption[CPath] 
{ \label{fig:CPath} 
Acquisition Camera and Common Path. a) CP (green) is mounted in the Nasmyth interface of telescope and the AC is on top (purple). b) General CP view, the light beam entrance is in the bottom center and the CAM selector send the light to the AC Unit. c) CAM selector with circles indicating the working position.}
\end{figure}

The backbone of the system is the CP, which receives the light from the telescope and distributes light to the sub-systems, see Fig.~\ref{fig:CPath}. 
The AC will receive the light from the CP through a $45^{\circ}$ mirror, called Selector Mirror, which has three permitted working positions: monitoring (takes the light that surrounds a $15$ asec hole), simulate an artificial star for calibration purposes (take the field surrounding a $0.5$ asec hole), and the imaging mode (full field). 
These functions will allow the AC to do photometry, to work in the acquisition of the target for spectrographs, for monitoring of spectrographs co-alignment, and also can be used as a guiding system if needed.

The SOXS project is nearing the commencement of operations, with expectations set for the instrument to become fully operational by 2025. In 2022, the Assembly, Integration, and Testing (AIT) for the AC was conducted. The AIT strategy adopted involved a two-stage approach. The initial stage focused on evaluating specific elements, comprising tests on the detector, assessment of filter curves, evaluation of alignment and performance of the lens tube with the detector, and examination of the light beam going through the AC Unit, intended for use in the experimental setup. These tests were conducted prior to the integration of the AC unit into the telescope simulator. Following this phase, the Assembly and Readiness Review (ARR) tests were conducted, the outcomes of which are detailed in this report.
The document is divided into four key sections: a thorough description of the instrument, a summary of the image quality tests used to assess the system, an outline of additional tests performed with the AC unit installed in the telescope simulator, and, ultimately, the conclusions derived from the complete testing process.

\section{AIT Overview}
\label{sec:AIT}

This section will describe the AC unit, including its characteristics, evaluation criteria, tolerance analysis, and the AIT strategy employed. The assembly and testing were conducted during the summers of 2022 and 2023 at the facilities of the Osservatorio Astronomico di Padova.

\subsection{Acquisition Camera}

\begin{figure} [ht]
\begin{center}
\begin{tabular}{c} 
\includegraphics[height=8.0cm]{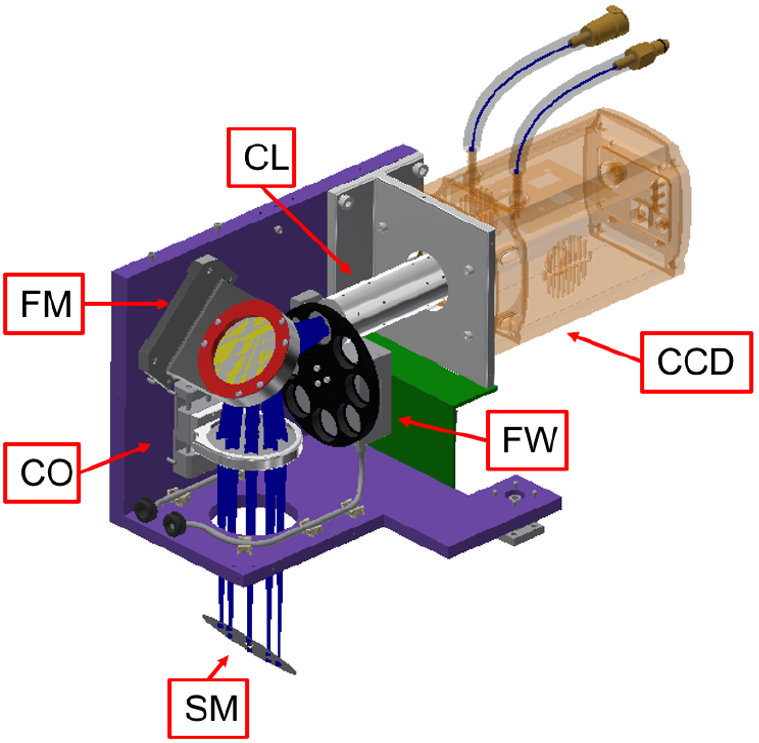}
\end{tabular}
\end{center}
\caption[ACUnit] 
{ \label{fig:ACUnit} 
AC Unit and its components.}
\end{figure}

The AC Unit consist of a Collimator Lens (CO), a folding mirror (FM), a filter wheel (FW), and a four lens camera, see Fig.~\ref{fig:ACUnit}.
The focusing mechanism is the CO, which is mounted in a linear stage.
The filter wheel also uses a rotating mechanism in order to choose between a broad-band filter set (ugrizY and V-Johnson). 
The detector is a $1024 x 1024$ Andor iKon-M 934 camera with a 13 micron pixel, and the control software was developed and validated by SOXS Consortium\cite{Andor,SOXS8:SPIE2018,SOXS3:SPIE2020}.   
The system includes a CAM selector, which will be mounted in the CP.
The CAM Selector carries a pellicle beam-splitter and the selector mirror (SM).
All the system is included in a structure made of 6061-T6 Aluminum and is anodized.

The AC Unit acts as a focal reducer, it reduces the incoming F/11 beam to a F/3.6.
Each pixel of the CCD will correspond to approximately $0.2$ asec, resulting in a $3.5$ amin unvignetted field of view.
The image quality specifications is such that over $80\%$ of the geometrically encircled energy is contained in two pixels for a central field of $6.5mm$ radius. 

\subsection{Tolerance Analysis}

After preparing all components for assembly, a new tolerance analysis was conducted to determine acceptable alignment limits, given the laser wavelength of 632 nm. This analysis examined three primary tolerance groups: (1) displacements along the optical axis (including separations between components and lens thicknesses), (2) lens radius of curvature, and (3) decentration and rotation of individual components and component groups. The sensitivity study indicated that stricter tolerances were required for elements within the optical tube. The Table ~\ref{tab:WOf} shows $10$ of the most critical areas needing attention. Based on these findings, tolerances for components outside the tube were set at $0.100$ mm for distances and $0.100$ degrees for rotation. For components inside the tube and the tube as a whole, tolerances were set at $0.050$ mm for distances and $0.050$ degrees for rotation.

\begin{table}[htbp]
\centering
\caption{\bf Worst Offenders.}
\begin{tabular}{p{1cm}|p{5cm}}
\hline
No. & Tolerance \\
\hline
$ 1$ & Camera Lens No.2 - Dec Y \\ 
$ 2$ & Tube - Rot Y \\ 
$ 3$ & Tube - Rot X \\ 
$ 4$ & Camera Lens No.2 - Dec X \\ 
$ 5$ & Camera Lens No.2 - Rot X \\ 
$ 6$ & Camera Lens No.1 - Rot X \\ 
$ 7$ & Camera Lens No.1 - Dec X \\ 
$ 8$ & Camera Lens No.2 - Surf 3 - Rad \\ 
$ 9$ & Camera Lens No.3 - Dec X \\ 
$10$ & Tube Spacer No.2 - Thick \\ 
\hline
\end{tabular}
 \label{tab:WOf} 
\end{table}

After establishing these tolerances, a Monte Carlo simulation with 1,000 trials was carried out to evaluate the feasibility of the arrangement. The design incorporates two compensators: the primary compensator is a linear stage with a $15$ mm stroke, on which the collimator lens is mounted. The second compensator is used exclusively during the alignment process and is a ring placed at the interface between the tube and the detector. These rings have fixed dimensions and come in four sizes: $0.050$, $0.100$, $0.250$, and $1.00$ mm. The criterion for evaluating the Monte Carlo simulation results is the distance containing $80\%$ of the energy of the Point Spread Function (PSF), or the $80\%$ Encircled Energy (EE).

\begin{figure} [ht]
\begin{center}
\begin{tabular}{c} 
\includegraphics[height=4.8cm]{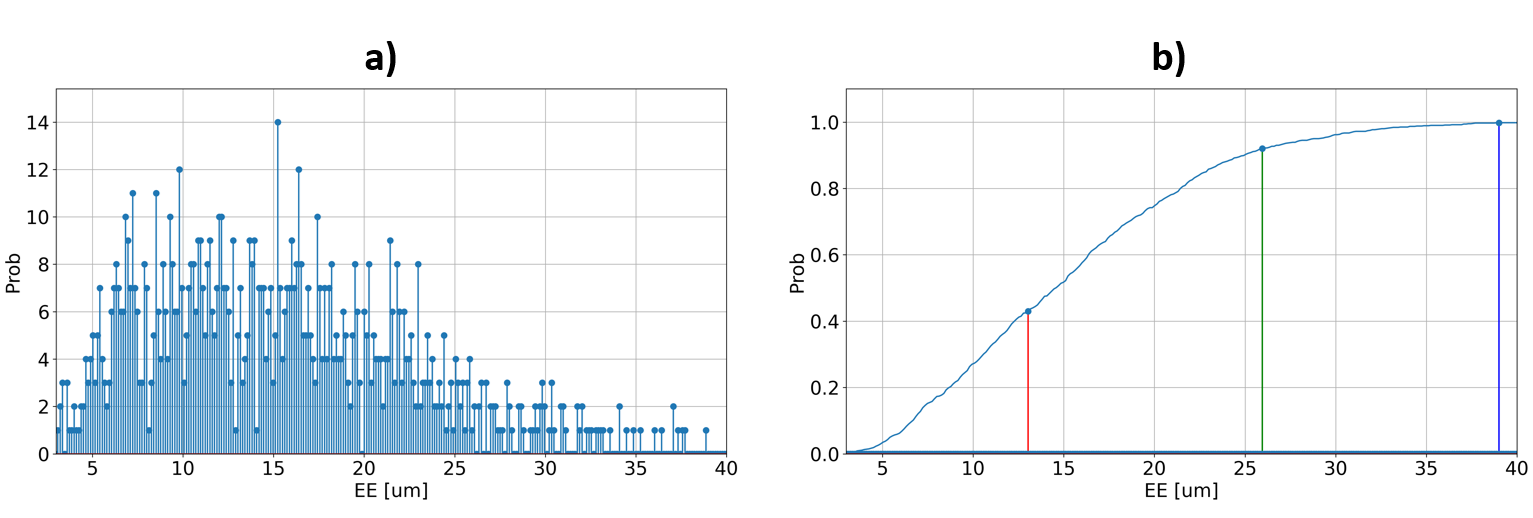}
\end{tabular}
\end{center}
\caption[TolA] 
{ \label{fig:TolA} 
Tolerance Analysis: $80\%$ of Encircled Energy (Diameter). a) Histogram of values per Monte Carlo case, and b) Cumulative Distribution Function (CDF). In plot b), the red, green, and blue lines represent the limits for 1, 2, and 3 pixels, respectively.}
\end{figure} 

Fig.~\ref{fig:TolA} presents a histogram of the Monte Carlo simulation results. The graph shows that in $92\%$ of cases, the energy is contained within two pixels, which meets the required specifications. The mean offset for the linear stage's compensator is $1.0$ mm from the center, while the average thickness for the compensator ring is $0.18$ mm. Based on these results, the assembly and alignment process proceeded with a maximum displacement tolerance of $100$ microns and a rotation tolerance of $0.1$ degrees for the overall arrangement. However, more stringent tolerances were applied during the insertion and alignment of lenses within the camera tube.

\subsection{Assembly and Readiness Review}

The concept and design of SOXS are modular, consisting of various sub-systems like the Common Path, UV-VIS and NIR Spectrograph, Calibration Unit, and the Acquisition Camera. Due to this modularity, each sub-system goes through an AIT phase at different partner locations. Afterward, an internal acceptance process is conducted, known as the ARR. The AIT is performed twice: once at INAF-OAPD under laboratory conditions, and once at the telescope site in La Silla. 

As part of the AIT phase, a detailed procedure was defined, outlining the steps to follow and a series of evaluations to be conducted. These evaluations were designed to ensure that the integration and alignment process was carried out correctly. The procedure to be followed consists of several steps, including:
\begin{itemize}
  \item Extender Cable Test.
  \item Camera Characterization.
  \item Filter transmission Curves.
  \item Camera's Optical Tube Alignment.
  \item Telescope Beam Simulator. 
  \item Image Quality Test. 
  \item Linear Stages Check. 
\end{itemize}
The last two steps, Image Quality Test and Linear Stages Check, are considered part of the ARR. 

\section{Image Quality Test}
\label{sec:IQT}

\begin{figure} [ht]
\begin{center}
\begin{tabular}{c} 
\includegraphics[height=4.8cm]{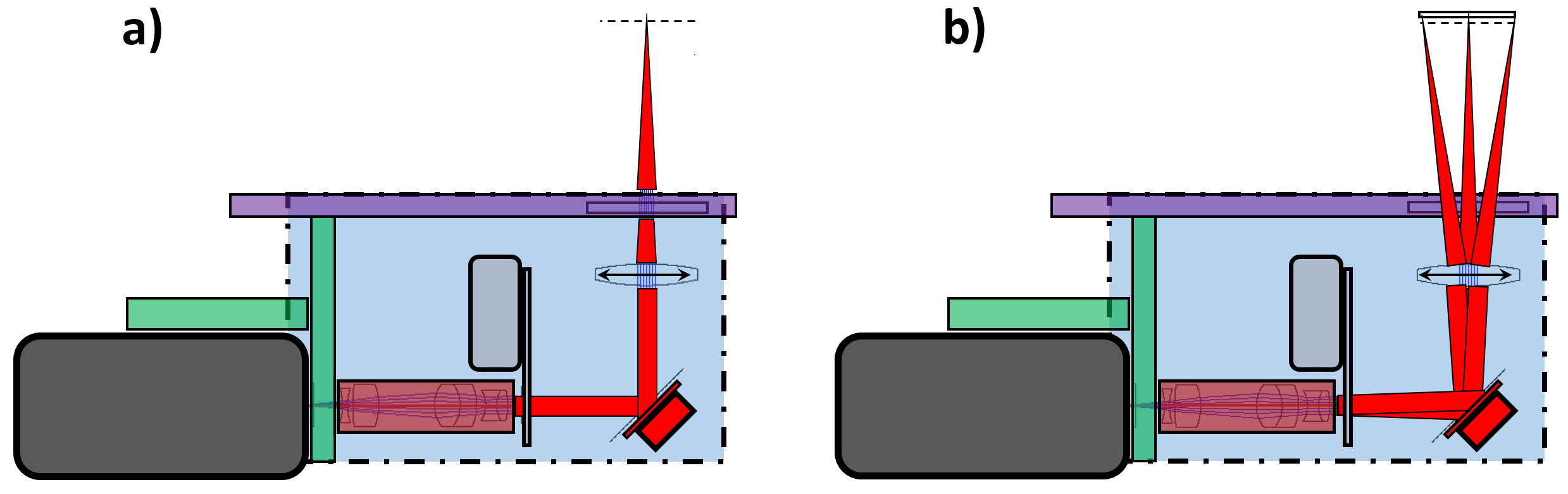}
\end{tabular}
\end{center}
\caption[OpB] 
{ \label{fig:OpBench} 
Experimental Setup for: a) Optical Alignment and b) Image Quality Test using Test Chart.}
\end{figure}

During the alignment process, characterized CCDs were used to trace the laser path along the array, while a Coordinate Measuring Machine (CMM) was employed to position components in accordance with specifications. The telescope's focal plane, with a light beam F/11, was evaluated by recording the generated PSF, after the plane's position had been previously defined using the CMM. Fig.~\ref{fig:OpBench}a shows the arrangement, illustrating the F/11 light beam diverging from the telescope's focal plane. 

Next, the Andor camera was used to begin capturing images, allowing for minor adjustments to achieve a PSF with rotational symmetry and a full width at half maximum (FWHM) of less than 2 pixels. This section focuses on evaluating image quality across the entire field using a specific filter. A test chart was employed to examine the resolution and distortion levels of the instrument, which was positioned in the telescope's focal plane, as depicted in Fig.~\ref{fig:OpBench}b. 

\subsection{Test Chart}

\begin{figure} [ht]
\begin{center}
\begin{tabular}{c} 
\includegraphics[height=5.2cm]{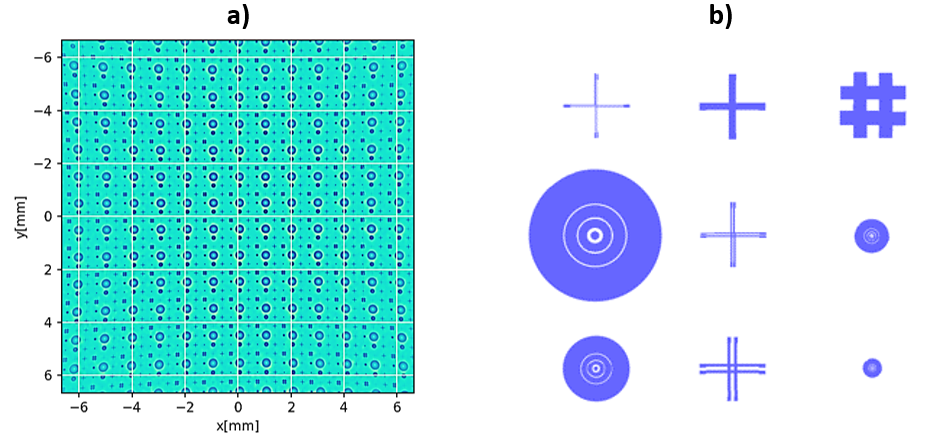}
\end{tabular}
\end{center}
\caption[TChart] 
{ \label{fig:TestCh} 
Concentric Circle and Distortion Grid Target: a) Picture taken with AC Unit, and b) Pattern Features.}
\end{figure}

The test chart used is a Thorlabs R3L3S5P\cite{TestCh}, with a substrate measuring $76.2$ x $72.6$ mm. The region containing the test patterns is $50.8$ x $50.8$ mm. The main grid comprises $17$ rows and $17$ columns of smaller grids, each containing nine patterns. These patterns include four concentric circles and five crosshairs of varying sizes, as shown in Figure \ref{fig:TestCh}b. The plate scale at the telescope's focal plane is $5.359$ arcseconds per millimeter (''/mm). This scale allows us to calculate the total area occupied by the test patterns in angular units. Given this scale, the area covered by the patterns is approximately $4.53$ x $4.53$ arcminutes. 

In the study conducted, two specific patterns were selected. For the resolution analysis, the largest concentric circular pattern was chosen, which is the first element in the second row of patterns. For the distortion analysis, the second-largest concentric circular pattern was used, which is the first element in the third row of patterns. For the analysis, the information from the internal rings of the circular patterns are not considered, since the outer contour of the circle forms the basis for the resolution analysis, and the center of mass of the pattern is calculated for the distortion analysis. As such, only the dimensions of the outer circles radius are required, which are $500$ and $250$ microns.

\subsection{Distortion Test}

Distortion is typically expressed as a percentage of the field height, commonly arising from aberrations near the edges of an image\cite{Dis1}. Generally, distortion within the range of $\pm 2$ to $3\%$ goes unnoticed in a vision system\cite{Dis2}. One common method to measure distortion in a camera-lens combination is the TV distortion method (ISO 9039), designed to analyze TV camera system\cite{Dis3}. This method uses a test chart with a rectangular grid of geometric patterns, such as the one shown in Fig.~\ref{fig:TestCh}b. The TV distortion method quantifies distortion by evaluating the curvature of a straight line from the image center to the corners. To calculate distortion, the degree of bending at the top edge of the image is measured. The ratio of this bending to the height of the image, multiplied by 100, yields the percentage of picture height distortion 
\begin{equation}
 D [\%] ={{\frac{\triangle H}{H}}\cdot 100}. 
\label{eq:di}
\end{equation}

To calculate the distortion associated with an optical system within a specific spectral range, defined by one of the filters, the process begins by applying a high-pass filter to clearly delineate the contour of the pattern being studied. A sample is taken from the image containing the selected pattern, and a convolution is performed on the entire image to find the centroid of these patterns. The centroid is then stored and later compared with the ideal geometry of the test chart. The following Fig.~\ref{fig:Disto}a illustrates an analysis of an image captured with the Z filter, displaying the measurement of every single pattern across the entire field (red dots), along with a continuous line calculated through numerical fitting (blue line). Using this approach, Fig.~\ref{fig:Disto}b presents the result of the fitted lines for each image taken with the different filters.

\begin{figure} [ht]
\begin{center}
\begin{tabular}{c} 
\includegraphics[height=5.4cm]{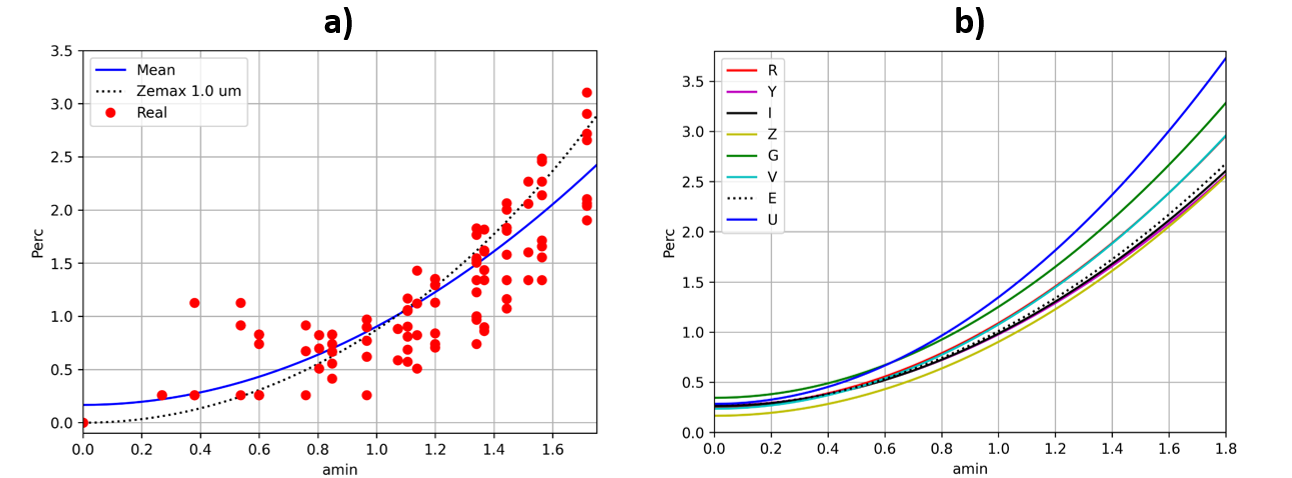}
\end{tabular}
\end{center}
\caption[Dis] 
{\label{fig:Disto} 
Distortion measurement: a) Actual measurements of each pattern with the Z filter along with the fitted line, and b) numerically fitted curves for all filters.}
\end{figure} 

\subsection{Resolution Test}

The next test to discuss is the sharpness test, which evaluates the system's capability to resolve fine details in a scene. The method used is a variant of the widely used slanted-edge spatial frequency response test, as outlined in the ISO 12233 standard\cite{Res1}. In this approach, samples from various horizontal or vertical scan lines are adjusted according to their estimated edge positions. These adjusted samples are then merged to create a new estimate of the edge transition with enhanced resolution. The spatial frequency response is derived by performing a Fourier transform on the first derivative of the edge transition function\cite{Res2}. The variation in this test lies in using the circular edge instead of an extended linear edge\cite{Res3}. Working with the circle's perimeter also allows us to increase measurement resolution. This method lets us use the same test chart that was used for distortion calculations, along with the same images. There are both advantages and disadvantages to this approach, which will be discussed at the end of this section, along with a comparison of the results against theoretical predictions derived from the optical design software (Zemax).

\begin{figure} [ht]
\begin{center}
\begin{tabular}{c} 
\includegraphics[height=5.4cm]{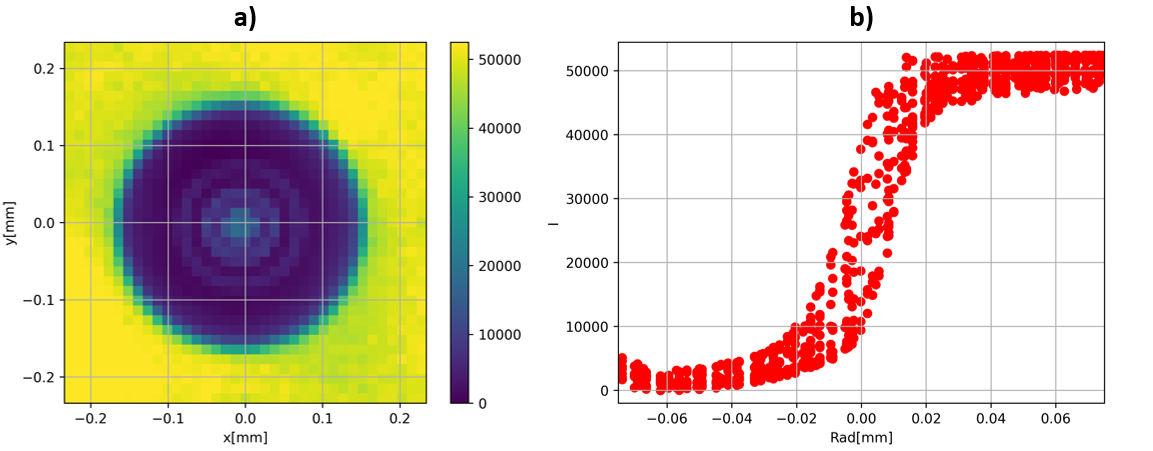}
\end{tabular}
\end{center}
\caption[Res1] 
{\label{fig:Res1} 
Edge Spread Function Calculation: a) selected pattern, and b) ESF of circle's boundary.}
\end{figure} 

The study begins by following the same instructions used in the distortion test. The selected pattern is used in a convolution procedure to obtain the centroids of the patterns across the image. Once the centroids are obtained, a segment of the image containing a pattern to be studied is selected, as shown in Fig.~\ref{fig:Res1}a. To calculate the circle's boundary, the coordinate space is in radial units. The study is limited to values within a radial range of $0.075$ to $0.240$ mm, these numbers are suitable for our application. 
Next, the image is analyzed pixel by pixel, and when a pixel falls within the specified region, its position and intensity are recorded in separate arrays. Figure \ref{fig:Res1}b shows a scatter plot of the results, with the ESF given as Radius [mm] versus Intensity [counts].
It is important to note that these recorded values are neither ordered nor uniformly spaced. The reference position of 0.0 mm, in Fig. \ref{fig:Res1}b, represents the distance from the ideal pattern radius which is $0.164$ mm.
The next step is to create two new arrays where the position vector has a constant separation. In this example, the sampling period achieved was 2.36 microns, a measurement at least five times smaller than the camera's pixel size, allowing us to conduct the Modulation Transfer Function (MTF) study described below.

\begin{figure} [ht]
\begin{center}
\begin{tabular}{c} 
\includegraphics[height=4.5cm]{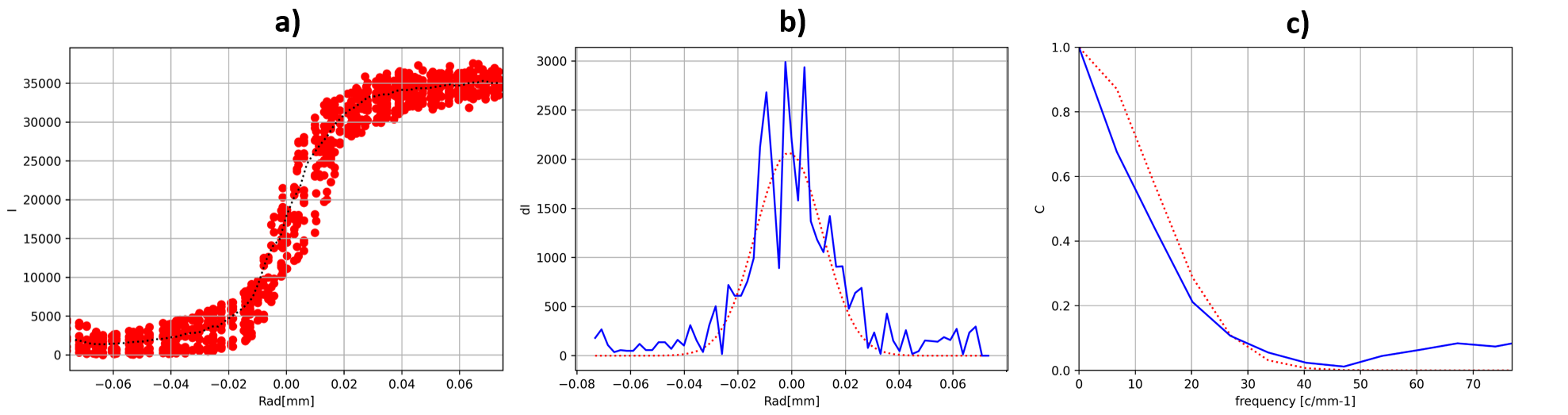}
\end{tabular}
\end{center}
\caption[Res2] 
{\label{fig:Res2} 
ESF, LSF and MTF Calculation.}
\end{figure} 

After obtaining the super-resolved version of the edge transition, the line-spread function (LSF) is estimated by differentiating the edge spread function (ESF). This differentiation can amplify high-frequency noise, so to mitigate this, a Gaussian curve is fitted to the LSF. The FWHM can then be derived directly from the Gaussian fit. Lastly, the MTF is calculated by applying a Fourier transform to the LSF curve. Fig.~\ref{fig:Res2} illustrates the process, showing the steps from the ESF to the LSF, and ultimately to the MTF.

\begin{figure} [ht]
\begin{center}
\begin{tabular}{c} 
\includegraphics[height=5.2cm]{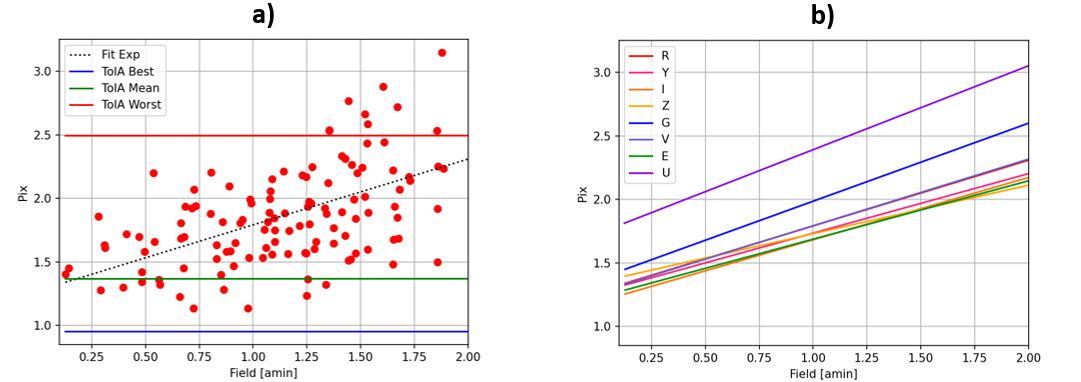}
\end{tabular}
\end{center}
\caption[Res3] 
{\label{fig:Res3} 
FWHM Results: a) Actual measurements with fitted lines derived from Monte Carlo cases, and b) Fitted lines from experimental data taken with each filter.}
\end{figure} 

As mentioned earlier, the FWHM can be derived from the LSF. By applying this procedure to each pattern throughout the image, it's possible to generate a map of FWHM values across the field of view. In Fig.~\ref{fig:Res3}a, the individual FWHM measurements taken from a specific image—using Filter R—are plotted. By fitting a line through these values, the dotted line is obtained. This fitted line can be compared with other lines representing the estimated FWHM from the best, mean, and worst designs according to the Monte Carlo analysis. The observation that $90\%$ of the measurements fall within the lines representing the best and worst cases from the Monte Carlo analysis provides confidence in the method's validity. Repeating this analysis with images taken for each filter results in Fig.~\ref{fig:Res3}b, illustrating the overall performance of the AC unit.

\section{AC Unit on Telescope Simulator}
\label{sec:AdTest}

When the image quality test was performed, the optimal focus positions for each filter were recorded, along with the best position for the filter wheel. After completing test, the AC Unit was installed in the telescope simulator and the numbers registered were checked. The filter wheel's position values were consistent, while the focal positions required slight adjustments to reach their final settings, see Table \ref{tab:Check}.

\begin{table}[htbp]
\centering
\caption{\bf Compensator and Filter Wheel Positions.}
\begin{tabular}{p{1cm}|p{3cm}|p{3cm}}
\hline
Filter & Linear Stage [mm] & Filter Wheel [Deg] \\
\hline
$ R $ & $  9.4$ & $239$\\ 
$ Y $ & $  5.8$ &  $14$\\ 
$ I $ & $  8.2$ & $149$\\ 
$ Z $ & $  6.5$ & $104$\\ 
$ G $ & $ 11.4$ &  $59$\\ 
$ V $ & $ 10.3$ & $194$\\ 
$ U $ & $ 12.7$ & $284$\\ 
$ E $ & $  6.7$ & $329$\\ 
\hline
\end{tabular}
 \label{tab:Check} 
\end{table}

\begin{figure} [ht]
\begin{center}
\begin{tabular}{c} 
\includegraphics[height=8.6cm]{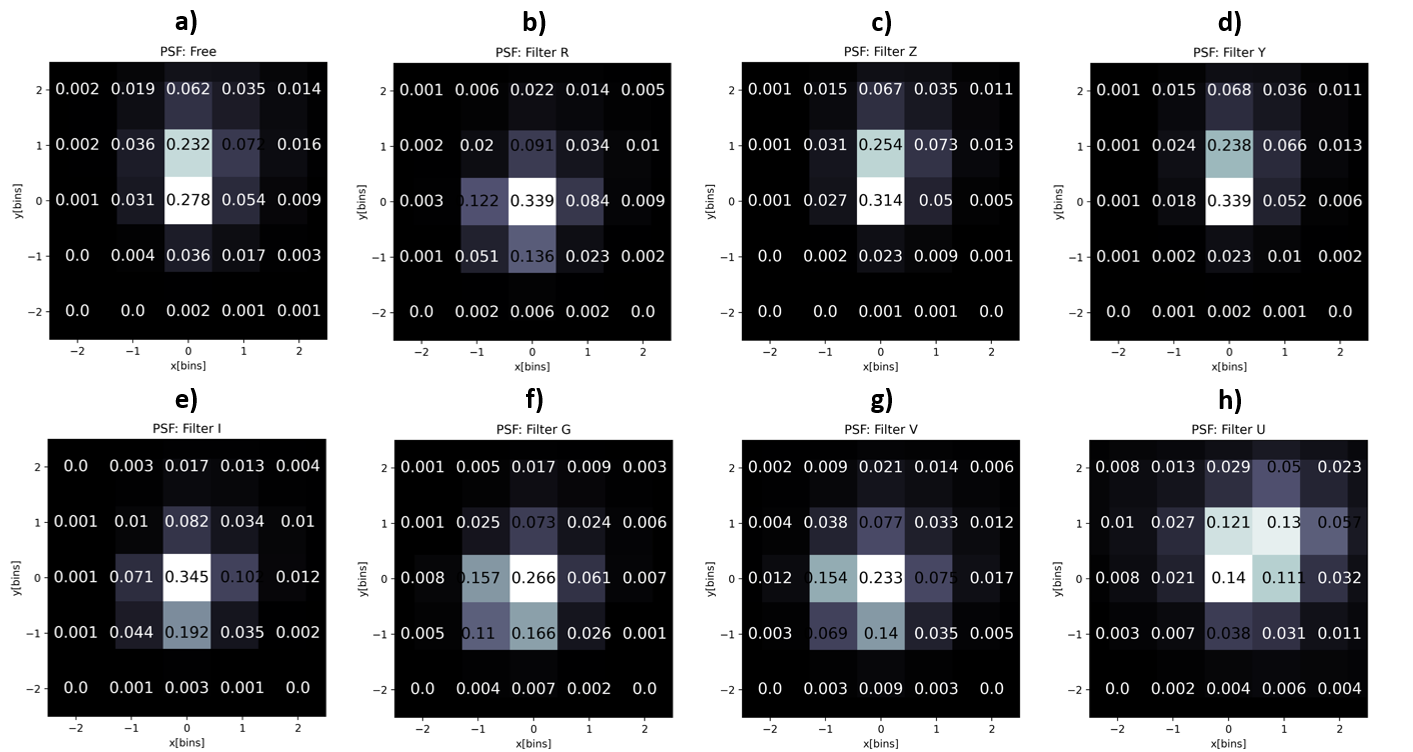}
\end{tabular}
\end{center}
\caption[psf] 
{\label{fig:psf} 
PSF Study of the AC Unit mounted on the Telescope Simulator. The study displays the PSFs captured with each filter slot: a) Free slot, b) Filter R, c) Filter Z, d) Filter Y, e) Filter I, f) Filter G, g) Filter 
 V, and h) Filter U.}
\end{figure}

In terms of optical performance, images were captured using the telescope simulator, which provides a $F/11$ beam with a halogen lamp. The following PSFs were obtained using the 7 available filters and the free position of the filter wheel, as shown in Fig.~\ref{fig:psf}. The graphs represent a $5x5$ matrix showing the information around the pixel with the maximum energy, with the total sum of the entire matrix equal to unity. When the light is concentrated in one or two pixels, the software used to capture the images has difficulty providing a reliable FWHM reading. This is why zooming in on a $5x5$ region of the total image helps visualize the energy distribution in greater detail. 
With this approach, the FWHM is typically contained within a radius of 2 pixels, indicating satisfactory performance. The results suggest that the system maintains a high level of precision and focus. Additionally, the flexure test revealed that the PSF center of mass shifted by an average of just 1 pixel during a full rotation of the flange. This limited displacement underscores the stability and robustness of the system's alignment under varying conditions.

\section{CONCLUSIONS}
\label{sec:sections}

The instrument is now fully integrated and operational. Final tests are being completed, and the required documentation is being prepared for transporting the instrument to the telescope\cite{SOXS2:SPIE2022,SOXS5:SPIE2022,SOXSD:SPIE2022,SOXS1:SPIE2020,SOXS8:SPIE2020,SOXSE:SPIE2020,SOXS6:SPIE2022}. The AC Unit was installed in the summer of 2022, with testing conducted in the summer of 2023. Although there were setbacks and mechanical issues with the optical components that complicated the alignment process, the problems were successfully resolved, and the system is now functioning as expected.

\section{ACKNOWLEDGMENT}
\label{sec:sections}

As part of the AC Unit-SOXS team, we'd like to extend our heartfelt thanks to the Padova work team for making us feel at home. We appreciate your support, guidance, and patience. Special thanks to Kalyan, Federico, Simone, Gigi, Davide, and Riccardo Claudi. We also want to acknowledge the Shark team, with whom we shared a workspace—they were equally welcoming. We're grateful to Mirko and Rosario from the SOXS electronics team for their help in resolving technical issues during system integration and testing. Thanks also to Matteo Alliverti and Matteo Munari for their support and guidance on aspects related to mechanical and optical design. Rafael Izazaga provided invaluable advice and assistance during the procurement of optical components, and Andrea Tozzi offered tremendous support throughout the integration process.

\bibliography{main} 
\bibliographystyle{spiebib} 

\end{document}